\documentstyle[preprint,aps]{revtex}

\begin{document}

\draft

\tighten

\title{On Adler-Bell-Jackiw Anomaly in 3-brane Scenario }

\author{ W.F. Chen${}^{1,2}$\renewcommand{\thefootnote}{\dagger}\thanks{
E-mail: w3chen@sciborg.uwaterloo.ca} and R.B.
Mann${}^{2}$\renewcommand{\thefootnote}{\ddagger}\thanks{ E-mail:
mann@avatar.uwaterloo.ca}}
\address{${}^1$ Department of Mathematics and Statistics,
University of Guelph\\
Guelph, Ontario, Canada N1G 2W1\\
${}^2$ Department of Physics, University of Waterloo\\
Waterloo, Ontario, Canada N2L 3G1}

\maketitle

\begin{abstract}
We investigate the ABJ anomaly in the framework of an effective
field theory for a 3-brane scenario and show that the contribution
from induced gravity on the brane depends on both the topological
structure of the bulk space-time and the embedding of the brane in
the bulk. This fact implies the existence of a non-trivial vacuum
structure of bulk quantum gravity. Furthermore, we argue that this
axial gravitational anomaly may not necessarily be cancelled by
choosing the matter content on the brane since it could be
considered as a possible effect from bulk quantum gravity.
\end{abstract}

\vspace{3ex}

\begin{flushleft}
{\bf 1.~Introduction}
\end{flushleft}

\vspace{2mm}

The idea that our observable space-time may be a
$(3+1)$-dimensional topological defect of some higher dimensional
quantum field theory has persisted over a number of years
\cite{akam,ruba}. In this scenario, the observed elementary
particles are the light particles trapped on the $(3+1)$
-dimensional defect and the Standard Model (SM), which is believed
to be the correct theory characterizing the interactions among
these elementary particles, appears as a low-energy effective
theory of a more fundamental theory in higher dimensions. The
exception is the graviton: it mediates the quantum gravitational
interaction, and it can propagate in the whole bulk due to the
equivalence between gravity and space-time geometry. Of course, it
might be possible that some other particles such as heavy fermions
-- beyond the reach of present accelerators -- also exist in the
bulk space-time. It is remarkable that such exotic yet simple
considerations can address some fundamental problems. For example,
it provides an alternative mechanism for solving the hierarchy
problem \cite{rand}, in contrast to those that modify the SM
itself such as technicolor and supersymmetric extensions. It also
gives a natural explanation as to why gravitational interactions
are much weaker than other forces \cite{arka}, and it even gives
an alternative means for addressing the cosmological problem
\cite{ruba,hold}.

This scenario has gained support from an application of
non-perturbative superstring theory. A physically realistic
example is that an $N=1$ $SU(5)$ supersymmetric gauge theory with
three generations of chiral matter fields can indeed come from one
sector of type-I string theory compactified on the $T^{6}/Z_{3}$
orientifold with five $D3$-branes placed at orientifold fixed
points \cite{lykk}. In general, a crude argument ignorant of the
specific brane configuration is the following: the 3-branes
provide a natural setting for the $(3+1)$-dimensional space-time,
the massless modes of the open string attached to the branes lead
to the observed gauge and matter fields, while the graviton comes
from the low-lying excited states of the closed string in bulk
space-time. The above situation is called the brane world scenario
\cite{tyeh}.

An explicit feature of the above brane scenario is that bulk
gravity is an essential ingredient. Consequently quantum gravity
in the bulk can affect physics on the brane. Concretely speaking,
in the case that the extra dimensional space is compact, in
addition to the massless graviton trapped on the brane, the
Kaluza-Klein (KK) states repsenting bulk gravity propagating in
the extra dimensions get involved in physical processes occurring
on the brane, some typical examples of which include the emission
of the KK graviton, the new scattering of SM particles from the
exchange of KK states and graviton and some higher order
corrections \cite{giud,hant,shiu}. In the case of large extra
dimensions, these new physical effects might be accessible to
testing via accelerator experiments in the near future
\cite{arka,shiu}. Some bold attempts have been made within this
framework to explain the recent measured deviation of the muon
anomalous magnetic moment from the SM prediction \cite{grae}.

However, as a reformed setting for describing elementary particle
interactions, particularly the role played by quantum gravity,
some dynamical features associated with gravity should be
reconsidered in this scenario. One typical problem is the famous
Adler-Bell-Jackiw (ABJ) anomaly for the axial vector current in a
chiral gauge theory defined on the brane. It is well known that
this anomaly can get a contribution in a background space-time
with non-trivial topology \cite{delb,eguc}, which is usually
called the axial gravitational anomaly in contrast to the pure
gravitational anomalies such as the Einstein and Lorentz anomalies
that arise in $D=4n+2$ dimensions \cite{alve,gins}.

The novel physical features are that bulk quantum gravity (which
at compact extra dimensions is effectively represented by the
dynamics of graviton, vector, and scalar fields and the
corresponding KK modes etc) becomes a dynamical field rather than
a static background, and that brane fluctuations can occur. It is
natural to ask whether or not these dynamical effects modify the
axial gravitational anomaly. Furthermore, if such contributions do
arise, what becomes of anomaly cancellation for a quantum field
theory defined on the brane? To our knowledge these issues have
not been explicitly addressed in the literature, and we consider
them in this paper.

\vspace{2mm}

\begin{flushleft}
{\bf 2.~A $U(1)$ chiral  gauge field model in 3-brane world}
\end{flushleft}

\vspace{2mm}

Let us start first from a simple model describing the low-energy
dynamics of a 3-brane, which includes the $D$-dimensional bulk
gravity and the chiral fermions confined on the brane interacting
with a $U(1)$ gauge field and bulk gravity through the induced
metric,

\begin{eqnarray}
S &=&S_{{\rm B.G.}}+S_{{\rm G}}+S_{{\rm F}}  \nonumber \\
&=&\frac{\kappa^{2}}{8}\int d^{D}X\sqrt{-G}R-\frac{1}{4}\int
d^{4}x\,e[X(x)]F_{\mu \nu }F^{\mu \nu }  \nonumber \\
&+&\int d^{4}x\,e[X(x)]\left\{e_{~a}^{\mu} \frac{1}{2}\left[
\left({\cal D} _{\mu }\overline{\psi }\right) i
\gamma^{a}-\overline{\psi }i\gamma^{a} {\cal D}_{\mu }\right]
\frac{1-\gamma_{5}}{2}\psi \right\} .  \label{model}
\end{eqnarray}
We consider only the one-flavour case here and one may add a
cosmological term for bulk gravity. The notation in the above
action is standard, $ X^{M}=(x^{\mu },y^{r})$, $M=0,\cdots ,D-1$,
the local coordinate of bulk space-time, $\mu =0,\cdots ,3$ and $r
=4,\cdots ,D-1$ being the coordinates of the brane and extra
dimensional space-time. To incorporate fermions on the brane, the
induced vierbein $e_{~\mu }^{a}[X(x)]$ and its inverse
$e_{~a}^{\mu }[X(x)]$ need some delicate consideration. As it is
well known, fermions on the 3-brane are the spinorial
representation of the local Lorentz group $SO(1,3)$. To guarantee
that this $SO(1,3)$ group is identical to the appropriate subgroup
of $SO(1,D-1)$, the local Lorentz group of bulk space-time, one
has to define \cite{sund}

\begin{eqnarray}
e_{~\mu }^{a}[X(x)]{\equiv }R_{~A}^{a}E_{~M}^{A}(X)B_{\mu }^{~M},
~~~B_{\mu}^{~M}{\equiv }\partial _{\mu }X^{M}(x),  \label{indve}
\end{eqnarray}
where $E_{~M}^{A}(X)$ is the vierbein corresponding to bulk
space-time metric $G_{MN}(X)=\eta_{AB}E_{~M}^{A}(X)E_{~N}^{B}(X)$,
$A=(a,m)$ is the bulk Lorentz index, and $a=0,\cdots ,3$,
$m=4,\cdots ,D-1$. $R(x)$ is actually an element of the bulk local
$SO(1,D-1)$ group depending only on the generator $J^{am}$
\cite{sund},

\[
R(x)=\exp [i\theta _{am}(x)J^{am}],
\]
and must satisfy

\begin{eqnarray}
R_{A}^{m}E_{~M}^{A}B_{\mu }^{~M}=0.  \label{cond}
\end{eqnarray}
The above two equations fix the requisite local Lorentz
transformation to define the correct vierbein induced on the brane
from the bulk metric. It has been shown that (\ref{indve}) and
(\ref{cond}) indeed lead to the induced metric \cite{sund}

\begin{eqnarray}
g_{\mu \nu }[X(x)]=\eta _{ab}e_{~\mu }^{a}[X(x)]e_{~\nu
}^{b}[X(x)]=G_{MN}(X)B_{\mu }^{~M}B_{\nu }^{~N} =\frac{\partial
X^{M}}{\partial x^{\mu }}\frac{\partial X^{M}}{\partial x^{\nu
}}G_{MN}(X),
\label{indme}
\end{eqnarray}
and hence $e[X(x)]=\sqrt{-g[X(x)]}$. Specifically the $SO(1,D-1)$
transformation in the bulk, $E_{~M}^{A}(X)\rightarrow
R_{~B}^{A}(X)E_{~M}^{B} $, will automatically lead to an $SO(1,3)$
rotation on the induced vierbein, $e_{~\mu }^{a}\rightarrow
r(x)_{~b}^{a}e_{~\mu}^{b} $. The fermions on the brane are just a
spinor representation of $r(x)_{~b}^{a}$. In this sense, the
spinor field on the brane is connected with the Lorentz symmetry
in the bulk. Of course, the other possibility is that we can start
directly from the representations of the Clifford group of the
bulk space-time (i.e. the covering group of $SO(1,D-1)$) and then
reduce it to the brane to get a spinor. However, since the
irreducible representations of the Clifford group in higher
dimensions depend heavily on the dimensionality, it seems to be
impossible to get a unique chiral gauge theory. Thus we define the
chiral fermions on the 3-brane from the representation of
$SO(1,3)$. For a $U(1)$ gauge field and induced gravitational
field on the brane, the operator ${\cal D}_{\mu }$ takes the usual
form

\[
{\cal D}_{\mu }=\partial _{\mu }-iYA_{\mu }
+\frac{1}{2}\omega_{\mu }^{ab}\sigma ^{ab},~~\sigma
^{ab}=\frac{1}{4}[\gamma ^{a},\gamma^{b}], ~~F_{\mu \nu
}=\partial_{\mu }A_{\nu }-\partial _{\nu}A_{\mu },
\]
where $Y$ is the charge carried by chiral fermions and for
simplicity we do not write out the gauge coupling explicitly. It
should be emphasized that the concrete form of the induced
vierbein (\ref{indve}) and metric (\ref{indme}) depends on the
dynamical behavior of the brane -- either moving in the bulk or
staying at a certain fixed point -- but the physics is equivalent
in these two cases.

It is easy to see that as for the usual 4-dimensional space-time
case, the fermionic part has a brane coordinate-dependent vector
and axial vector gauge transformation,

\begin{eqnarray}
\psi (x)\rightarrow \exp [iY\theta (x)]\psi (x), ~~\overline{\psi
} (x)\rightarrow \overline{\psi }(x)\exp [-iY\theta (x)],~~A_{\mu
}(x)\rightarrow A_{\mu }(x)+\partial _{\mu }\theta (x),
\label{vett}
\end{eqnarray}
and

\begin{eqnarray}
\psi (x)\rightarrow \exp [iY\gamma _{5}\vartheta (x)]\psi
(x),~~\overline{ \psi }(x)\rightarrow \overline{\psi }(x)\exp
[iY\gamma_{5}\vartheta (x)],~~A_{\mu }(x)\rightarrow A_{\mu
}(x)+\partial _{\mu }\vartheta (x), \label{avett}
\end{eqnarray}
as well as the reparametrization invariance of the brane, the
infinitesimal version being \cite{gins}

\begin{eqnarray}
&&x^{\mu }\rightarrow x^{\mu }-\xi ^{\mu }(x),~~ \delta
e_{~\mu}^{a}(x)=e_{~\nu }^{a}\nabla _{\mu }\xi ^{\nu }+\xi ^{\nu
}\nabla_{\nu}e_{~\mu }^{a},~~\delta e=\partial _{\mu
}(\xi^{\mu}e),
\nonumber \\
&&\delta \omega _{\mu }^{~ab}=\xi ^{\nu }\partial _{\nu }\omega
_{\mu }^{~ab}+\omega _{\nu }^{~ab}\partial _{\mu }\xi ^{\nu
},~~\delta \psi =\xi^{\mu }\partial _{\mu }\psi ,~~\delta
\overline{\psi }=\xi^{\mu }\partial_{\mu }\overline{\psi }.
\label{rept}
\end{eqnarray}
At the quantum level, the axial vector gauge symmetry cannot be
simultaneously upheld with the vector gauge symmetry and the
reparametrization invariance of the brane, and so becomes
anomalous.

\vspace{2mm}

\begin{flushleft}
{\bf 3.~ Quantization of 3-brane world and ABJ anomaly }
\end{flushleft}

\vspace{2mm}

Before turning to the axial gravitational anomaly, we briefly look
at the quantization of the model. Even bypassing the
renormalizability problem of bulk quantum gravity in the above
effective field theory model, we still have no way to completely
quantize the system with such a matter distribution. Since the
brane world is somehow an effective field theory description,
there are two perspectives one can adopt toward the quantization
of such a physical system. The first perspective is that widely
adopted in the literature: if the extra dimensions are compact,
the bulk gravitational field in general admits an expansion in
terms of the orthonormal modes living in the extra space and the
KK modes on the world volume of 3-brane. In this framework, one
can get an effective theory describing the interaction between the
matter fields on the brane and the KK modes after integrating out
the extra dimensions. The effects of the bulk gravitational field
on the brane can then be detected by studying the quantization of
this effective action. The second perspective is quite formal, but
is universal to any brane world models regardless of what the
extra space is like, either compact or having infinite size like
the second class Randall-Sundrum model \cite{rs2}. In this
perspective (the one we shall adopt), one first quantizes the
field theory on the brane, obtaining a quantum effective action
relevant to bulk gravity, then subsequently considers the
quantization of bulk gravity. These two versions of quantization
of the 3-brane world should (at least qualitatively) lead to
consistent results for quantum phenomena in the brane world when
the extra dimensions are compact.

The second viewpoint shall shape our interpretation of the axial
gravitational anomaly in 3-brane world. Note that the dynamics in
the bulk is invariant under both diffeomorphisms and $SO(1,D-1)$
transformations, while on the brane, the theory has the
reparametrization invariance, local $SO(1,3)$ symmetry and various
gauge symmetries at the classical level. One must choose gauge
conditions to eliminate the redundant degrees of freedom connected
with these symmetries of bulk gravity; gauge-fixing and the
relevant ghost terms shall arise as usual for the gauge theory.
The quantum theory of this system can be formally written out
using the path integral

\begin{eqnarray}
Z &=&\int \prod_{M,N}{\cal D}H_{MN}(X)\prod_{\mu }{\cal
D}\overline{A}_{\mu }(x){\cal D}\overline{\psi }(x){\cal D}\psi
(x)\exp {i\left[ S+\cdots \right]
}  \nonumber \\
&=&\int \prod_{M,N}{\cal D}H_{MN}(X)\left\{ \exp \,i\left[
\frac{\kappa ^{2} }{8}\int d^{D}X\sqrt{-G}R+\cdots \right] \int
\prod_{\mu }{\cal D}\overline{A}_{\mu }(x){\cal D} \overline{\psi
}(x){\cal D}\psi (x)\right.
\nonumber \\
&&\left. \times \exp \left[ \int d^{4}x\,e[X(x)]\left( e_{~a}^{\mu
}[X(x)] \overline{\psi }i\gamma ^{a}\frac{1-\gamma ^{5}}{2}{\cal
D}_{\mu }\psi - \frac{1}{4}F_{\mu \nu }F^{\mu \nu }+\cdots \right)
\right] \right\}
\nonumber \\
&=&\int \prod_{M,N}{\cal D}H_{MN}(X)\left\{ \exp \,i
\left[\frac{\kappa ^{2}}{8}\int d^{D}X\sqrt{-G}R+\cdots \right]
\right.  \nonumber \\
&&\times \prod_{\mu }{\cal D}\overline{A}_{\mu }(x)\exp \left[
i\int d^{4}x\,e[X(x)]\left( -\frac{1}{4}F_{\mu \nu }F^{\mu \nu
}+\cdots \right)\right]  \nonumber \\
&&\times \left. \det \left[ ie(X)\,e_{~a}^{\mu }(X)\gamma
^{a}\frac{1-\gamma^{5}}{2}{\cal D}_{\mu }\right] \right\}  \nonumber \\
&=&\int \prod_{M,N}{\cal D}H_{MN}(X)\left\{ \exp \,i\left[
\frac{\kappa ^{2}
}{8}\int d^{D}X\sqrt{-G}R+\cdots \right] \right.  \nonumber \\
&&\times \left. \prod_{\mu }{\cal D}\overline{A}_{\mu }(x) \exp
\left[ i\int d^{4}x\,e[X(x)]\left( -\frac{1}{4}F_{\mu \nu }F^{\mu
\nu }+\cdots \right) \right] \,\exp \left( iW[A,e(X)]\right)
\right\} ,  \label{quse}
\end{eqnarray}
where we write $G_{MN}(X)=G_{MN}^{(0)}(X)+H_{MN}(X)$, $A_{\mu
}(x)=A_{\mu }^{(0)}(x)+\overline{A}_{\mu }(x)$ i.e., we adopt the
general view that the bulk graviton is a spin-2 quantum field over
certain background space-time $G_{MN}^{(0)}$ and that there exists
a vacuum configuration $A_{\mu }^{(0)}$ for the gauge field. These
may or may not be trivial ($G_{MN}^{(0)}=\eta_{MN} $, $A_{\mu
}^{(0)}=0$), depending on the case under consideration. The
ellipses denote the gauge-fixing and ghost terms for the
diffeomorphism invariance of bulk space-time and the gauge
symmetry on the brane as well as a possible cosmological term for
bulk gravity. In particular, we write out the explicit dependence
of the induced metric (or vierbein) on the bulk coordinate in
order to show that the quantum effective action of the field
theory is intimately related to brane dynamics in the bulk
space-time. With the setting (\ref{quse}) for the quantization of
brane world, we are now able to discuss the possible anomalies for
the model (\ref{model}). Like the usual 4-dimensional chiral gauge
theory in the gravitational and gauge field background, the
effective action $W[A,e(X)]$ cannot remain invariant under all the
transformations given by (\ref{vett}), (\ref{avett}) and
(\ref{rept} ). According to the definition, the effective action
$W[A,e(X)]$ of the model (\ref{model}) can be formally written as
the sum of the Green functions of the current operators\footnote{
It should be emphasized that at this stage both gauge and
gravitational fields are purely background fields rather than the
quantum ones.},

\begin{eqnarray}
&&W[A,e(X)]\sim \sum \int \left[ \prod_{i}\left(
d^{4}x_{i}e[X(x_{i})]\right) A_{\mu _{i}}(x_{i})\prod_{j}\left(
d^{4}y_{j}e[X(y_{j})]\right) A_{\nu _{j}}(y_{j})\right.  \nonumber \\
&\times &\left. \prod_{k}\left( d^{4}z_{k}e[X(z_{k})]\right)
g_{\lambda _{k}\rho _{k}}(z_{k})\left\langle
\prod_{i}\widehat{J}^{5\mu_{i}}(x_{i})\prod_{j}
\widehat{J}^{\nu_{j}}(y_{j})\prod_{k} \widehat{T}_{{\rm
(F)}}^{\lambda _{k}\rho _{k}}(z_{k})\right\rangle _{{\rm
connected}}\right] ,
\end{eqnarray}
where the axial vector current $J_{\mu }^{5}$, vector current
$J_{\mu }$ and the fermionic part of energy-momentum tensor
$T_{(F)\mu \nu }$ are, respectively,

\begin{eqnarray}
J_{\mu }(x) &=&Y\overline{\psi }(x)\gamma _{\mu }\psi (x),  \nonumber \\
J_{\mu }^{5}(x) &=&Y\overline{\psi }(x)\gamma _{\mu }\gamma
_{5}\psi ,(x),
\nonumber \\
T_{{\rm (F)}\,\mu \nu } &=&\frac{2}{e[X(x)]}\frac{\delta S_{{\rm
F}}}{\delta g^{\mu \nu }}=i\left[ \overline{\psi }\left( \gamma
_{\mu }{\cal D}_{\nu }+\gamma _{\nu }{\cal D}_{\mu }\right)
\frac{1-\gamma ^{5}}{2}\psi -g_{\mu \nu }\overline{\psi }\gamma
^{\lambda }{\cal D}_{\lambda }\frac{1-\gamma^{5} }{2}\psi \right].
\end{eqnarray}
It is well known that the contribution to $W[A,e(X)])$ from the
Green functions $\langle \widehat{J}_{\mu }^{5}(x)\widehat{J}_{\mu
}(y)\widehat{J} _{\nu }(z)\rangle $ and $\langle \widehat{J}_{\mu
}^{5}(x)\widehat{T}_{(F)\nu \rho }(y)\widehat{T}_{(F)\lambda
\sigma }(z)\rangle$ can not make the axial vector gauge symmetry
compatible with both the vector gauge symmetry and the
reparametrization invariance of the brane \footnote{ A possible
arising of Lorentz anomaly is ignored.},

\begin{eqnarray}
i\delta W[A,e(X)] &=&-\frac{i}{16}\int
d^{4}xd^{4}yd^{4}ze[X(x)]e[X(y)]e[Z(z)]  \nonumber \\
&&\times \left\{ \vartheta (x)\left[ A_{\nu }(y)A_{\rho
}(z)\left\langle\partial _{\mu }\widehat{J}^{5\mu
}(x)\widehat{J}^{\nu }(y)
\widehat{J}^{\rho }(z)\right\rangle \right. \right.  \nonumber \\
&&\left. -g_{\nu \lambda }(y)g_{\rho \sigma }(z)\left\langle
\partial _{\mu} \widehat{J}^{5\mu }(x)
\widehat{T}_{(F)}^{\nu \lambda}(y)
\widehat{T}_{(F)}^{\rho \sigma }(z)\right\rangle \right]  \nonumber \\
&&+2A_{\mu }(x)A_{\nu }(y)\theta (z)\left\langle \widehat{J}^{5\mu
}(x) \widehat{J}^{\nu }(y)\partial _{\rho }\widehat{J}^{\rho
}(z)\right\rangle
\nonumber \\
&&\left. -2A_{\mu }(x)g_{\nu \lambda }(y)\xi _{\sigma
}(z)\left\langle \widehat{J}^{5\mu }(x)\widehat{T}_{(F)}^{\nu
\lambda }(y)\nabla _{\rho }
\widehat{T}_{(F)}^{\rho \sigma }(z)\right\rangle \right\}  \nonumber \\
&=&i\int d^{4}x\,e[X(x)]\left[ \frac{1}{2}\vartheta (x)\partial
_{\mu }\left\langle \widehat{J}^{5\mu }\right\rangle
-\frac{1}{2}\theta (x)\partial _{\mu }\left\langle
\widehat{J}^{\mu }\right\rangle -\xi _{\nu }(x)\nabla _{\mu
}\left\langle \widehat{T}_{(F)}^{\mu \nu }\right\rangle \right] ,
\end{eqnarray}
where $\nabla _{\mu }$ being the covariant derivative defined with
respect to Levi-Civita symbol of the induced metric. To carry out
a concrete calculation of the anomaly, one usually makes a
decomposition $g_{\mu\nu}=\eta _{\mu \nu }+h_{\mu \nu } $ and
consider linearized gravity when $|h_{\mu \nu }|\ll 1$
\cite{alve}. As a consequence, one has $g^{\mu \nu }=\eta ^{\mu
\nu }-h^{\mu \nu }$, and $ e[X(x)]=1+\frac{1}{2}\eta ^{\mu \nu
}h_{\mu \nu }$. The fermonic part of the classical action
(\ref{model}) can be approximately written as

\begin{eqnarray}
S_{{\rm (F)}} &=&-\frac{1}{2}\int d^{4}x\,e[X(x)]
g^{\mu\nu}T_{{\rm (F)}\mu\nu }  \nonumber \\
&=&\int d^{4}x\left[ i\overline{\psi }\eta ^{\mu \nu } \gamma
_{\mu }{\cal D} _{\nu }\frac{1-\gamma ^{5}}{2}\psi
+\frac{1}{2}h^{\mu \nu }T_{(F)\mu \nu } \right] ,
\end{eqnarray}
where the subsidiary conditions $\eta ^{\mu \nu }\partial _{\mu
}h_{\nu \rho }(x)=\eta ^{\mu \nu }h_{\mu \nu }(x)=0$ are used.
With the requirement of preserving vector gauge symmetry and the
general covariance on the brane, i.e., choosing $\partial _{\mu
}\langle \widehat{J}^{\mu }\rangle =\nabla _{\mu }\langle
\widehat{T}^{\mu \nu }\rangle =0$, we have the anomaly for the
axial vector current. It is a long-standing result that a direct
calculation to the triangle and seagull diagrams of above
three-point function gives \cite{delb,eguc}

\begin{eqnarray}
\partial _{\mu }\left\langle\widehat{J}^{5\mu }\right\rangle
=\frac{Y^{3}}{16\pi ^{2}} \epsilon^{\mu \nu \lambda \rho }
F_{\mu\nu }F_{\lambda \rho } + \frac{Y}{ 384\pi ^{2}} \epsilon
^{\lambda \rho \sigma \delta }R_{~\nu \lambda \rho
}^{\mu}R_{~\mu\sigma \delta }^{\nu }.  \label{chian}
\end{eqnarray}

\vspace{2mm}

\begin{flushleft}
{\bf 4.~ Dependence of axial gravitational anomaly on embedding of
3-brane in a factorizable  bulk space-time}
\end{flushleft}

\vspace{2mm}

In the 3-brane scenario, the first term of the chiral anomaly
given in (\ref{chian}) comes from the instanton configuration of
the gauge field confined on the brane, which is identical to the
usual case since it is independent of the background space-time
metric. The second term, contributed from the induced
gravitational instanton background \cite{eguc}, should be relevant
to the classical Euclidean configuration of the bulk gravitational
field. To show this connection explicitly, let us recall briefly
the submanifold theory in Riemannian Geometry \cite{kono}. For a
$d$-dimensional submanifold $M$ of a $D$-dimensional Riemannian
manifold ${\cal N}$ with a relation between their local
coordinates, $X^{M}=X^{M}(x^{\mu })$, $M=0,1,\cdots ,D-1$ , $\mu
=0,1,\cdots ,d-1$, one can define a quantity of rank $d$,
$B_{\mu}^{~M}{\equiv }\partial X^{M}/\partial x^{\mu }$, which
plays a role of both a covariant vector in the submianifold and a
contravariant vector in the bulk manifold. $B_{\mu }^{~M}$
connects the local differential structure of the submanifold with
that of the bulk manifold. The basis of their tangent spaces are
related by

\[
\frac{\partial }{\partial x^{\mu }}=B_{\mu }^{~M}\frac{\partial
}{\partial X^{M}},
\]
and hence there exists a relation between the induced metric on
the manifold and the bulk metric, $g_{\mu \nu }=G_{MN}B_{\mu
}^{~M}B_{\nu }^{~N}$. A tangent space of the bulk manifold at a
generic point can be decomposed into an orthogonal direct sum of
the tangent space of the submanifold with a $n-m$-dimensional
vector space equipped with orthonormal basis, $N_{r}$,

\[
\left\langle \frac{\partial }{\partial x^{\mu
}},N_{r}\right\rangle =0,~~r=d-1,\cdots ,D-1.
\]
This $D-d$-dimensional vector space is called normal space of the
submanifold $M$. There exist the following relations according to
the definition,

\[
G_{PQ}B_{\mu
}^{~P}N_{r}^{~Q}=0,~~G_{PQ}N_{r}^{~P}N_{s}^{~Q}=\delta _{rs},
\]
where $N_{r}^{~P}$ are the components of $N_{r}$ in bulk
space-time. Assuming that $(B_{\mu }^{~P},N_{r}^{~Q})$ have the
inverse $(\overline{B}_{P}^{~~\mu },\overline{N}_{Q}^{~~r})$ with
respect to the bulk manifold indices $P$, $Q$, i.e.,

\begin{eqnarray}
&&B_{\mu }^{~P}\overline{B}_{P}^{~~\nu }=\delta _{\mu }^{~\nu
},~~~N_{r}^{~P}
\overline{N}_{P}^{~~s}=\delta _{r}^{~s},  \nonumber \\
&&B_{\mu
}^{~P}\overline{N}_{P}^{~~r}=N_{r}^{~P}\overline{B}_{P}^{~~\mu
}=0,
\end{eqnarray}
one can easily derive the following equations,

\begin{eqnarray}
&&\overline{B}_{P}^{~~\nu }g_{\nu \mu }=B_{\mu
}^{~Q}G_{QP},~~\overline{B}_{P}^{~~\mu }=G_{PQ}\overline{B}_{~\nu
}^{Q}g^{\nu \mu },~~B_{\mu}^{~P}=G^{PQ}B_{Q}^{~~\nu }g_{\nu \mu },
\nonumber \\
&&g^{\mu \nu }=G^{PQ}\overline{B}_{P}^{~~\mu
}\overline{B}_{Q}^{~~\nu },~~ \overline{N}_{P}^{~r}=\delta
^{rs}N_{s}^{~Q}G_{QP},~~N_{r}^{~P}=G^{QP}
\overline{N}_{Q}^{~~s}\delta _{sr},
\end{eqnarray}
and

\begin{eqnarray}
&&\overline{B}_{Q}^{~~\mu }B_{\mu
}^{~P}+\overline{N}_{Q}^{~~r}N_{r}^{~P}= \delta _{Q}^{~P},~~g_{\mu
\nu }\overline{B}_{P}^{~~\mu }\overline{B} _{Q}^{~~\nu }+\delta
_{rs}\overline{N}_{P}^{~~r}\overline{N}
_{Q}^{~~s}=G_{PQ},  \nonumber \\
&&\frac{1}{(D-d)!}\overline{N}_{P_{1}}^{~r_{1}}\cdots
\overline{N}_{P_{D-d}}^{~~~r_{D-d}}\epsilon _{r_{1}\cdots
r_{D-d}}=\frac{1}{d!}\frac{1} {\sqrt{-G}}\epsilon ^{\mu _{1}\cdots
\mu _{d}}B_{\mu _{1}}^{~Q_{1}}\cdots B_{\mu _{d}}^{~Q_{d}}\epsilon
_{P_{1}\cdots P_{D-d}Q_{1}\cdots Q_{d}}.
\end{eqnarray}
In addition, the covariant derivative $\nabla _{\mu }B_{\nu
}^{~P}$ for fixed indices $\mu $, $\nu $ is actually a normal
vector of the submanifold $M$,

\begin{eqnarray}
\nabla _{\mu }B_{\nu }^{~P}&=& \partial _{\nu }B_{\mu
}^{~P}-\Gamma _{\mu \nu }^{~~\lambda }B_{\lambda }^{~P}+B_{\mu
}^{~R}B_{\nu }^{~S}\Gamma _{RS}^{~~~M}  \nonumber
\\
&=&\left( \partial _{\nu }B_{\mu }^{~Q}+B_{\mu }^{~M}B_{\nu
}^{~N}\Gamma _{MN}^{~~~~Q}\right)
\overline{N}_{Q}^{~~r}N_{r}^{~P}\nonumber\\
&{\equiv }& K_{\mu \nu }^{~~~P} =K_{\nu \mu }^{~~P}=K_{\mu \nu
}^{~~r}N_{r}^{~P},  \nonumber \\
K_{\mu \nu }^{~~r} &{\equiv }&\left( \partial _{\nu }B_{\mu
}^{~Q}+B_{\mu }^{~M}B_{\nu }^{~N}\Gamma _{MN}^{~~~~Q}\right)
\overline{N}_{Q}^{~~r},
\nonumber \\
&&G_{PQ}B_{\mu }^{~P}K_{\lambda \rho }^{~~Q}=0,
\end{eqnarray}
and the covariant derivative of $N_{r}^{~P}$ is given by the
Weingarten formula,

\begin{eqnarray}
\nabla _{\mu }N_{r}^{~P} &=&-\delta _{rs} \left( g^{\lambda
\nu}K_{\mu\lambda }^{~~s}B_{\nu }^{~P}-L_{\mu }^{~st}N_{t}^{~P}\right),
\nonumber \\
L_{~s\mu }^{r} &{\equiv }&N_{s}^{~P}\nabla _{\mu
}\overline{N}_{P}^{~~r}
\end{eqnarray}
In above equations, $\Gamma _{\mu \nu }^{~~\lambda }$ and $\Gamma
_{PQ}^{~~~M}$ are the Christoffel symbols for the submanifold and
the bulk manifold, respectively, and they have the following
relation,

\[
\Gamma _{\mu \nu }^{~~\lambda }=\overline{B}_{P}^{~~\lambda
}\left( \partial _{\mu }B_{\nu }^{~P}+B_{\mu }^{~R}B_{\nu
}^{~S}\Gamma _{RS}^{~~~P}\right).
\]
With above equations one can derive the Gauss, Codacci and Ricci
equations,

\begin{eqnarray}
R_{~\nu \lambda \rho }^{\mu } &=&B_{\nu }^{~N}B_{\lambda
}^{~P}B_{\rho }^{~Q}R_{~NPQ}^{M}\overline{B}_{M}^{~~\mu }+K_{\rho
~R}^{~\mu }K_{\lambda
\nu }^{~~R}-K_{\lambda ~R}^{~\mu }K_{\rho \nu }^{~~R},  \nonumber \\
N_{r}^{~N}B_{\rho }^{~Q}B_{\lambda }^{~P}R_{~NPQ}^{M}
\overline{B}_{M}^{~~\mu } &=&\left( \nabla _{\lambda }K_{\rho
~r}^{~\mu }-\nabla _{\rho }K_{\lambda ~r}^{~\mu }\right) +\left(
L_{\rho r}^{~~s}K_{\lambda ~s}^{~\mu }-L_{\lambda
r}^{~~s}K_{\rho ~s}^{~\mu }\right) ,  \nonumber \\
B_{\mu }^{~M}B_{\nu }^{~N}N_{r}^{~P}B_{s}^{~Q}R_{MNPQ} &=&K_{\mu
~r}^{~\sigma }K_{\sigma \nu s}-K_{\nu ~r}^{~\sigma }K_{\sigma \mu
s}+\nabla_{\mu }L_{\nu rs}-\nabla _{\nu }L_{\mu rs}  \nonumber \\
&&-\left( L_{\mu r}^{~~t}L_{\nu ts}-L_{\nu r}^{~~t}L_{\mu
ts}\right) .
\end{eqnarray}
The Gauss equation shows how the Riemannian curvature tensor of
the submanifold is related to the bulk one, and the quantity
$K_{\rho ~R}^{~\mu }K_{\lambda \nu }^{~~R}-K_{\lambda ~R}^{~\mu
}K_{\rho \nu }^{~~R}$ is the extrinsic curvature tensor of the
submanifold in the bulk manifold, which is completely composed the
normal vectors of the submanifold. The Codacci and Ricci equations
further gives how the Riemannian curvature tensor of bulk manifold
is projected into the tangent and normal spaces at a generic point
of the submanifold.

With above equations specializing to 4-dimensional submanifold
case, we can rewrite the gravitational part of the chiral anomaly
in terms of the bulk Riemannian tensor and the quantities
characterizing the embedding of 3-brane in bulk space-time,

\begin{eqnarray}
\partial _{\mu }\left\langle \widehat{J}^{5\mu }\right\rangle
=\frac{Y}{384\pi ^{2}}\left( A_{1}+A_{2}+A_{3}\right) ,
\label{anbulk}
\end{eqnarray}
 where

\begin{eqnarray}
A_{1} &{\equiv }&\left( \overline{B}_{M_{1}}^{~~\mu }B_{\mu
}^{~N_{2}}\right) \left( \overline{B}_{M_{2}}^{~~\nu }B_{\nu
}^{~N_{1}}\right) \epsilon ^{\lambda \rho \sigma \delta
}B_{\lambda}^{~P_{1}}B_{\rho }^{~Q_{1}}B_{\sigma
}^{~P_{2}}B_{\delta
}^{~Q_{2}}R_{~~N_{1}P_{1}Q_{1}}^{M_{1}}R_{~~N_{2}P_{2}Q_{2}}^{M_{2}}
\nonumber \\
&=&\sqrt{-G}\frac{1}{[(D-4)!]^{2}}\epsilon ^{R_{1}\cdots
R_{D-4}P_{1}Q_{1}P_{2}Q_{2}}\epsilon _{r_{1} \cdots
r_{D-4}}\overline{N}
_{R_{1}}^{~~r_{1}}\cdots \overline{N}_{R_{D-4}}^{~~r_{D-4}}  \nonumber \\
&\times &\left[
R_{~~QP_{2}Q_{2}}^{P}R_{~~PP_{1}Q_{1}}^{Q}-2R_{~~~QP_{2}Q_{2}}^{M_{2}}
R_{~~N_{1}P_{1}Q_{1}}^{Q}
\overline{N}_{M_{2}}^{~~~r}N_{r}^{~N_{1}}\right.   \nonumber \\
&&\left. +\overline{N}_{M_{1}}^{~~r}N_{r}^{~N_{2}}\overline{N}
_{M_{2}}^{~~~s}N_{s}^{~N_{1}}R_{~~N_{2}P_{2}Q_{2}}^{M_{2}}
R_{~~N_{1}P_{1}Q_{1}}^{M_{1}}
\right] ;  \nonumber \\
A_{2} &\equiv &4\epsilon ^{\lambda \rho \sigma \delta } K_{\rho
~R}^{~\mu }K_{\lambda \nu }^{~~R}\overline{B}_{M}^{~\nu }B_{\mu
}^{~N}B_{\sigma}^{~P}B_{\delta }^{~Q}R_{~~NPQ}^{M};  \nonumber \\
A_{3} &=&4\epsilon ^{\lambda \rho \sigma \delta }K_{\rho ~R}^{~\mu
}K_{\lambda \nu }^{~~R}K_{\delta ~S}^{~\nu }K_{\sigma \mu
}^{~~~S}. \label{conc}
\end{eqnarray}
The geometric meaning of above three terms are obvious. $A_{1}$
represents the topological invariant constructed from the
projection of the bulk Riemanian curvature tensor into the tangent
space of the 3-brane world volume, which can be equivalently
described in terms of the bulk curvature tensor together with the
normal vectors of the 3-brane world volume. $A_{2}$ gives a
topological invariant constructed from extrinsic curvature tensor
and the projection of bulk Riemannian curvature tensor into the
tangent space, while $A_{3}$ is a topological invariant built
purely from the extrinsic curvature tensor. In spite of the
induced metric $g_{\mu \nu }$ of the submanifold being a
projection of the bulk metric into the tangent space, the
Riemannian curvature tensor corresponding to the induced metric is
not identical to the projection of bulk curvature tensor into the
tangent space, since according to the Gauss equation there exists
an extrinsic curvature tensor relevant to the normal space. Thus
for a submanifold it is equivocal to speak of the topological
meaning of the Pontrjagin class constructed from the Riemannian
curvature tensor corresponding to the induced metric. It is
necessary to pull the induced Riemannian curvature to the bulk
manifold and discuss the topological meaning of the corresponding
Pontrjagin class.

Now we convert above geometric objects into physics using the
equivalence of Riemmanian geometry and gravity.
Eqs.\,(\ref{anbulk}) and (\ref{conc}) imply immediately that the
axial gravitational anomaly observed in bulk space-time depends on
both the topological structure of the bulk and the immersion of
the brane, i.e., how the brane is geometrically located in bulk
space-time. One may think that this conclusion does not make
sense, since naive considerations suggest that since the axial
vector current is confined to the brane, the gravitational anomaly
should reside only in the topology of the brane. This would be
true if the 3-brane were not embedded in a higher dimensional
space-time and the fields $X^{M}(x)$ describing the position of
the brane were not dynamical fields. The dependence of the axial
gravitational anomaly on the dynamics of the brane and bulk
gravity can be further explained as follows. In general, there are
two possibilities for the 3-brane in the bulk: the 3-brane either
moves freely or sits at a fixed point of the extra dimension(s)
\cite{giud}. In the former case, it is redundant to describe the
position of the brane in terms of the bulk space-time coordinate
$X^{M}(x)$. One can eliminate this redundancy by choosing

\[
X^{\mu }(x)=x^{\mu },~~X^{r}(x)=\xi ^{r }(x), ~~\mu=0,\cdots,3,~~
r=4,\cdots,D.
\]
The induced metric contains explicitly fields $\xi ^{\alpha }(x)$
defined on the brane, which are called branons and describe the
fluctuations of a 3-brane in bulk space-time \cite{giud}. In the
latter case, there exists \cite{giud}

\[
X^{M}(x)=x^{\mu }\delta _{\mu }^{~M}.
\]
The induced metric coincides with the bulk metric in the
directions that the brane extends, $g_{\mu \nu }=G_{MN}\delta
_{~\mu }^{M}\delta _{~\nu }^{N}$. However, in this case, the
translation invariance of the extra dimensions is broken. The
branons $\xi (x)$ will still arise as a Goldstone fields
corresponding to the breaking of translational symmetry in the
directions of extra dimensions. The dynamical effect of this kind
of Goldstone boson was discussed in Ref.\cite{doba}.

\vspace{2mm}

\begin{flushleft}
{\bf 5.~Brane world in non-factorizable bulk space-time: RS1
model}
\end{flushleft}

\vspace{2mm}

In the above, we considered the brane world to be of a type that
the bulk space-time is factorizable, i.e., the position of the
brane can be completely determined by the bulk coordinates as
functions of the brane coordinates. However, there exist some
brane worlds not separable from the extra dimension in the sense
that the induced metric may depend on the coordinate of the extra
dimension, which turns out to have more useful applications in
particle physics phenomenology than the decomposable case
\cite{rand}. In the following, we discuss this case by considering
a concrete model --- the first Randall-Sundrum model (RS1)
\cite{rand} --- and show explicitly how the axial gravitational
anomaly presents in this kind of brane scenario.

RS1 consists of two parallel 3-branes of opposite tension in the
5-dimensional bulk space-time, which turns out to be a slice of
$AdS_{5}$ space. The extra dimension is a line segment, the
orbifold $S^{1}/Z_{2}$ parametrized by the coordinate $y\in
\lbrack -\pi ,\pi ]$. The two 3-branes localize at the orbifold
fixed points of $Z_{2}$, $y=0,\pi $, respectively. Contrary to the
usual case with an extra dimension, the world-volume of the
3-brane and the extra dimension is non-factorizable. Despite
having two 3-brane worlds, only the one with negative tension
(localized at $y=\pi $ -- called the visible brane) is the brane
supporting the SM model; the other one (called the hidden brane)
is just a necessary set-up to produce a warp factor in the 3-brane
metric and generate the hierarchy. Since the extra space of the
RS1 model is compact, it is convenient to use the first version
stated above to discuss quantization. The classical solution to
the bulk Einstein equation is the warped metric \cite{rand}

\[
ds^{2}=e^{-2kr_{c}|y|}\eta _{\mu \nu }dx^{\mu }dx^{\nu
}+r_{c}^{2}dy^{2}.
\]
where $k$ is the $AdS_{5}$ curvature, $M_{{\rm Pl}}$ the
4-dimensional Planck scale and $r_{c}$ the radius of the extra
dimension $S^{1}$. The bulk quantum gravity is described by the
quantum fluctuation around the above warped metric \footnote{ The
excitation of the modulus field (or radion) is not considered here
and a mechanism is assumed to keep the metric component of the
extra dimension frozen at $r_{c}$ \cite{gowi}.} \cite{rand}:

\begin{eqnarray}
ds^{2} &=&G_{\mu \nu }(x,y)dx^{\mu }dx^{\nu }+r_{c}^{2}dy^{2},
\nonumber \\
G_{\mu \nu }(x,y) &=&e^{-2kr_{c}|y|}\left[ \eta _{\mu \nu }
+h_{\mu\nu }(x,y) \right]
\end{eqnarray}
Since the extra dimension is compact, the quantum fluctuations
admit an orthonormal mode expansion \cite{davo},

\begin{eqnarray}
&&h_{\mu \nu }(x,y)=\sum_{n=0}^{\infty }h_{\mu \nu
}^{(n)}(x)\frac{\chi_{n}(y)}{\sqrt{kr_{c}}},  \nonumber \\
&&\int_{-\pi }^{\pi }dye^{-2kr_{c}y}\chi _{m}(y)\chi
_{n}(y)=\delta _{mn},
\end{eqnarray}
where $h_{\mu \nu }^{(m)}$ and $h_{\mu \nu }^{(n)}$ are the
graviton and the K-K modes respectively. In addition, they satisfy
the gauge conditions $\eta^{\mu \nu }\partial _{\mu }h_{\nu \rho
}^{(n)}(x)=0$ and $\eta ^{\mu \nu }h_{\mu \nu }^{(n)}(x)=0$. The
four-dimensional effective Lagrangian density describing the
interaction of KK modes with the matter fields on the visible
3-brane is \cite{davo}

\begin{eqnarray}
{\cal L} &=&-\frac{1}{M^{3/2}}T^{\mu \nu }(x)h_{\mu \nu
}(x,y)|_{y=\pi }
\nonumber \\
&=&-\frac{1}{M_{{\rm Pl}}}T_{(F)}^{\mu \nu }\left[ h_{\mu \nu
}^{(0)}+e^{kr_{c}\pi }\sum_{n=1}^{\infty }h_{\mu \nu
}^{(n)}\right], \label{eff}
\end{eqnarray}
where $M$ is the 5-dimensional Planck scale and is related to its
4-dimensional counterpart on the visible 3-brane through the
relation $M_{{\rm Pl}}^{2}=(1-e^{-2kr_{c}\pi })M^{3}/k$
\cite{rand}. In situations where the energy-momentum tensor of the
matter fields consists of chiral fermions as the model
(\ref{model}), if we consider gravity at linearized level, the
effective Lagrangian (\ref{eff}) implies that the axial
gravitational anomaly takes the following form,

\begin{eqnarray}
\partial _{\mu }\left\langle \widehat{J}^{5\mu}
\right\rangle_{{\rm GRA }} &\sim &\frac{1}{M^{3}} \epsilon ^{\mu
\nu \lambda \rho }R_{~\delta \mu \nu }^{\sigma }(x,y)R_{~\sigma
\lambda \rho }^{\delta }(x,y)|_{y=\pi }  \nonumber
\\
&=&\epsilon ^{\mu \nu \lambda \rho }\left[ \frac{1}{M_{{\rm
pl}}^{2}} R_{~~~~\delta \mu \nu }^{(0)\sigma }R_{~~~~\sigma
\lambda \rho }^{(0)\delta }+\frac{2e^{kr_{c}\pi }}{M_{{\rm
pl}}^{2}}R_{~~~~\delta \mu \nu }^{(0)\sigma }\sum_{n=1}^{\infty
}R_{~~~~\sigma \lambda \rho }^{(n)\delta }\right.
\nonumber \\
&&\left. +\frac{e^{2kr_{c}\pi }}{M_{{\rm
pl}}^{2}}\sum_{n,m=1}^{\infty }R_{~~~~\delta \mu \nu }^{(n)\sigma
}R_{~~~~\sigma \lambda \rho }^{(m)\delta }\right] ,
\label{j5current}
\end{eqnarray}
where $R_{~~~~\nu \lambda \rho }^{(0)\mu }$ and $R_{~~~~\nu
\lambda \rho }^{(n)\mu }$ are the Riemannian curvatures
corresponding to $h_{\mu \nu }^{(0)}$ and $h_{\mu \nu }^{(n)}$
respectively. We have employed the result that for linearized
gravity there exists the expansion

\[
R_{~\delta \mu \nu }^{\sigma }(x,y)=\sum_{n=0}^{\infty
}R_{~~~~\delta \mu \nu }^{(n)\sigma }\frac{\chi
^{n}(y)}{\sqrt{kr_{c}}}.
\]

The meaning of Eq.\,(\ref{j5current}) needs some explanation. The
chiral anomaly gets contributions from both the graviton on the
brane and the KK modes. As it is well known, like the chiral
anomaly in the gauge field background, the topological origin of
the axial gravitational anomaly is the fermionic zero modes in the
gravitational instanton background \cite{eguc}, a Euclidean
solution to the Einstein equation with (anti-)self-dual Riemannian
tensor. It is just the existence of this kind of Euclidean
configuration in the gravitational field that leads to the
difference of the chiral fermonic zero modes and thereby generates
the axial gravitational anomaly. The existence of a gravitational
instanton implies that the corresponding quantum gravity theory
must have a non-trivial vacuum structure because an instanton lies
between two distinct vacua and produces a tunneling effect between
them. Hence we naturally relate the $R^{(0)} \widetilde{R}^{(0)}$
term in Eq.\thinspace (\ref{j5current}) to the gravitational
instanton background on the 3-brane. The $R^{(0)}\widetilde{R}
^{(n)}$ and $R^{(0)}\widetilde{R}^{(n)}$ terms are the
contributions from the infinite tower of KK modes. Although one
can similarly explain them in terms of an infinite tower of
gravitational instanton-like objects, they are actually connected
to the non-factorizable feature of the bulk space-time and to the
topological structure of the extra space.

Let us make this point clear by recalling the origin of KK modes.
In general, the appearance of gravitational KK modes comes from a
decomposition of the bulk space-time into a physical space-time
and a compact extra space. Any bulk field $H(X)$, $X=(x,y)$, has a
mode expansion, $H(x,y)=\sum_{n}h^{(n)}(x)\chi _{n}(y)$. The
topological structure of the extra space determines the
orthonormal modes $\chi _{n}(y)$ and thereby the KK modes. In
particular, the isometry group of the extra dimensions becomes a
global symmetry of the effective theory defined in the physical
space-time and even a dynamical symmetry if this global symmetry
can be gauged. In this sense, the physical property of KK modes
depends intimately on the topological structure of the extra
dimension. This is consistent with our inference that the axial
gravitational anomaly is relevant to the bulk space-time. An
explicit investigation of how the geometry of large but compact
extra dimensions in a factorizable bulk space-time affects the
field theory on the 3-brane through KK mode-emission was carried
out in Ref.\cite{fred}, and it was shown there that the low-level
KK modes can distinguish the topology of the extra space, whereas
the high energy modes seem not to be sensitive to it. However, RS1
model is special in the sense that the bulk space-time is not
factorizable. The induced metric is thereby identical to the
warped metric \cite{shir} and has a dependence on the coordinate
of extra dimension. Consequently, the induced metric admits a KK
modes expansion which leads to the anomaly of the form
(\ref{j5current}). From a phenomenological consideration, the
result (\ref{j5current}) implies that a physical process
associated with the above chiral anomaly can receive contributions
from every levels of KK modes. In particular,
Eq.\,(\ref{j5current}) shows that the contributions from the
graviton and KK modes are graded by the energy scales $M_{{\rm
pl}}$ and $e^{-kr_{c}\pi }M_{{\rm pl}}$ . With an appropriate
choice on the size of the extra dimension, one can make
$e^{-kr_{c}\pi }M_{{\rm pl}}$ be at the order of the weak scale.
Thus it is possible to probe the existence of such an extra
dimension with the physical process described by anomalous
diagrams.

Can we express the axial gravitational anomaly in terms of the
bulk Riemannian tensor and the quantities describing the embedding
of 3-brane and understand the corresponding topological meaning in
the same way as the case of decomposable bulk space-time? In
general, it is no clear how to rigorously define a submanifold
theory for this case. The reason is that the induced metric is
dependent on the  extra dimension(s), $g_{\mu \nu }=g_{\mu \nu
}(x,y)$. This implies that the bulk coordinates characterizing the
embedding of the brane depend on the extra dimension,
$X^{M}=X^{M}(x,y)$. As a consequence, this may imply that the rank
of $B_{\mu }^{~M}(x,y)$ is less than $d$, making it impossible to
define the normal vectors and normal space of the submanifold.
However, in some special cases such as the $RS1$ model, the brane
is fixed in certain point of the extra dimension and the normal
vectors can be well defined. Thus we can write the Riemannian
curvature tensors of the brane, which include all the curvatures
for the zero and KK modes in the case that the extra dimensions
are compact, in terms of the corresponding bulk Riemannian
curvature tensors and extrinsic curvature tensors as the
decomposable bulk space-time case.

\vspace{2mm}

\begin{flushleft}
{\bf 6.~ Anomaly cancellation and possible origin from bulk
quantum gravity}
\end{flushleft}

\vspace{2mm}

Turning next to the anomaly cancellation problem, the gauge field
contribution, i.e., $F\widetilde{F}$ must be cancelled as usual.
Thus there is no inconsistency for the $SU(2)_{L}\times U(1)_{Y}$
SM defined on the brane, since the requirement
$\sum_{i}Y_{i}^{3}=0$ for the cancellation of the anomaly
contributed from the gauge field is equivalent to the condition
$\sum_{i}Y_{i}=0$ for the gravitational anomaly cancellation for
the SM fermions, the index $i$ denoting the flavours of the
fermionic particles. This can be easily verified with the
Gellman-Nishijima formula, $Y=2(Q-T_{3}) $, and the fact that the
electromagnetic current is anomaly free as a vector current
\cite{cheng}, $Q$ and $T_{3}$ being the electric charge and the
remaining generator of $SU(2)_{L}$ after spontaneous breaking.
However, there exist some extensions of the SM that are free of
gauge field part but not of gravitational field contributions
\cite{alve}. If this occurs, it is not necessary to cancel the
gravitational anomaly part of the brane world, in contrast to the
usual case. One straightforward observation is that the
topological number density $\epsilon ^{\mu \nu \lambda \rho
}R_{~\delta \mu \nu }^{\sigma }R_{~\sigma \lambda \rho }^{\delta
}$ can be written as a total divergence \cite{alve},

\begin{eqnarray}
&& \epsilon ^{\mu \nu \lambda \rho }R_{~\delta \mu \nu }^{\sigma
}R_{~\sigma\lambda \rho }^{\delta }=\nabla _{\mu }K^{\mu },  \nonumber \\
&& K^\mu =4
\epsilon^{\mu\nu\lambda\rho}\left(\Gamma^{\sigma}_{~\nu\delta}\partial_
\lambda \Gamma^{\delta}_{~\rho\sigma}+\frac{2}{3}
\Gamma^\sigma_{~\nu\delta} \Gamma^\delta_{~\lambda\alpha}
\Gamma^\alpha_{~\rho\sigma} \right), \label{kmu}
\end{eqnarray}
$\Gamma^{\lambda}_{~\mu\nu}$ being the Christoffel symbols with
respect to the induced metric on the brane. Thus one can make a
shift on the quantum effective action

\begin{equation}
\overline{W}[A,e(X)]=W[A,e(X)]+\frac{1}{384\pi ^{2}}\int
d^{4}x\,e[X(x)]A_{\mu }(x)K^{\mu }[X(x)].  \label{effact1}
\end{equation}
$\overline{W}[A,e(X)]$ is invariant under the gauge transformation
$\delta A_{\mu }=\partial _{\mu }\vartheta (x)$ but violates the
reparametrization invariance given in (\ref{rept}), since $K^{\mu
}$ is not invariant under general covariant coordinate
transformations (\ref{rept}). However, the breaking of general
covariance on the brane is not a serious problem, since this can
be considered as an effect of bulk quantum gravity. This can be
easily observed as follows. Under the infinitesimal coordinate
transformation given in (\ref{rept}), the induced metric vary as
$\delta g_{\mu \nu }=\nabla _{\mu }\xi _{\nu }+\nabla _{\nu }\xi
_{\mu }$, we have

\begin{eqnarray}
\delta \overline{W}[A,e(X)] &=&\int d^{4}x\,e[X(x)]\frac{\delta
\overline{W} [A,e]}{\delta g^{\mu \nu }(x)}\left( \nabla ^{\mu
}\xi ^{\nu }+\nabla
^{\nu}\xi ^{\mu }\right)  \nonumber \\
&=&-\int d^{4}xe[X(x)]\xi ^{\nu }\nabla ^{\mu }\langle T_{\mu \nu
}\rangle. \label{noco}
\end{eqnarray}
Eq.\,(\ref{noco}) relates directly the origin of the anomalous
general covariance on the brane to $K^{\mu }$. However, as shown
in (\ref{kmu}), $K_{\mu }$ is a functional of the induced metric
on the brane. General covariance of the bulk space-time remains
unaffected by the redefinition (\ref{noco}) of the effective
action since the induced metric is a scalar with respect to
general coordinate transformations of the bulk space-time. There
is no reason to exclude the possibility that the $K_{\mu }$ term
can be generated from bulk quantum gravity since it is completely
consistent with the bulk space-time symmetry. Of course, the
concrete physical process of generating such a term is not clear
since we know little about bulk quantum gravity. In brane-world
scenarios, the general covariance of bulk space-time is the most
fundamental symmetry. Reparametrization invariance on the brane,
represented by the induced metric, is dominated by the general
covariance of bulk space-time and the way in which the brane is
embedded in the bulk. Based on above considerations we can impute
the anomalous breaking of general covariance on the brane to the
quantum effects of bulk gravity.

The above arguments reveal a remarkable feature of brane world
scenarios: viewed from the brane perspective, quantum bulk
gravitational effects can in principle render a chiral gauge
theory on the brane quantum mechanically inconsistent, breaking
either general covariance or chiral gauge symmetry. Roughly
speaking, quantum fluctuations of bulk gravity described by $
G_{MN}(X)$ can affect the induced metric $g_{\mu \nu }(x)$ through
the relation (\ref{indme}) by generating the term $K^{\mu}$, which
couples with the dynamical gauge field $A_{\mu }(x)$ on the brane.
The quantum effective action for the brane system is then
described by $\overline{W}\left[ A,e(X) \right] $, which is
invariant under general coordinate transformation of bulk
space-time, but not with respect to the reparametrization of the
brane. To preserve general covariance on 3-brane, one must
redefine the quantum effective action as

\begin{eqnarray}
W[A,e(X)]=\overline{W}[A,e(X)]-\frac{1}{384\pi ^{2}}\int
d^{4}x\,e[X(x)]A_{\mu }(x)K^{\mu }[X(x)].  \label{eff2}
\end{eqnarray}
It should be emphasized that viewed from the bulk perspective the
above redefinition of the quantum effective action is a finite
renormalization, and both $W$ and $\overline{W}$ characterize the
same bulk physics. However, $W$ and $\overline{W}$ describe
distinct quantum phenomena for the brane system: for a physical
process represented by $W[A,e(X)]$, general covariance is
preserved but the chiral symmetry is violated, while the converse
takes place for $\overline{W}$.

There are two perspectives one can adopt based on the preceding
considerations. The first is that both general covariance and
chiral symmetry must be preserved on the brane. In this case the
second term in Eq.\,(\ref{effact1}) must vanish, constraining the
possible form of the bulk quantum theory of gravity. However, an
alternative perspective is to regard Eq. (\ref{effact1}) as
providing a means to probe quantum gravity in the bulk. The term
$\displaystyle\int d^{4}xe[X(x)]A_{\mu }(x)K^{\mu }[X(x)]$ will be
deduced from observation of physical processes on the 3-brane: the
axial vector (or chiral) current on the brane in a gravitational
background definitely receives a gravitational anomalous breaking
either in chiral symmetry or in general covariance of the brane
system. Keeping the brane theory gauge anomaly-free (as indicated
by present-day experiments), observations of apparent quantum
inconsistencies as described by Eqs.\thinspace (\ref{effact1}) or
(\ref{eff2}) could be regarded as providing signature effects of
bulk quantum gravity. Since the term responsible for breakdown of
general covariance of the brane system is actually allowed by the
symmetry of the full bulk theory, such a viewpoint cannot be ruled
out.

How viable is such a perspective? Although a concrete physical
process leading to $K^{\mu }[X(x)]$ is not yet clear, the
dynamical phenomenon of anomaly inflow (relevant to defects such
as strings, domain walls and $p$-branes) \cite{caha,nacu} provides
further support for this viewpoint. If defects like strings and
domain walls exist in a given space-time, there will in general be
chiral fermionic zero modes trapped on the defect, even if one
starts with a vector type gauge theory (i.e., the matter fields
are Dirac fermions). According to the Atiyah-Singer index theorem
\cite{egha}, a chiral anomaly localized on the defect must arise.
This means that the gauge charges (in case of a gauge anomaly) or
the energy and momentum (in case of a gravitational anomaly)
carried by the chiral fermionic zero modes trapped on the defect
are not conserved. Equivalently the fermionic determinant
contributed from the chiral fermionic zero modes localized on the
defect is not gauge invariant (or not generally covariant).
However the full bulk theory must be anomaly-free since one began
with a vector gauge theory. Thus the gauge charges (for a gauge
anomaly) or energy and momentum (for gravitational anomaly)
carried by the massive fermionic modes away from the defects will
also not be conserved and must ``flow'' into the defect to
compensate for the non-conservation due to the anomaly localized
on the defect. In other words the fermionic determinant
contributed by the massive fermions away from the defect must be
also not gauge invariant, counteracting the anomaly in the
fermionic zero mode defect determinant to ensure chiral gauge
symmetry.

From a brane-world perspective this dynamical mechanism could be
called anomaly ``outflow ''. It entails a loss of energy and
momentum conservation on the brane because the anomaly permits
energy and momentum transfer from brane to bulk (or vice-versa)
via the anomaly. In the scenario we have considered, the chiral
fermions on the 3-brane are put in arbitrarily by hand and the
bulk theory is pure gravity. However, we can still envision an
extension of the bulk theory that can provide the requisite
anomaly inflow mechanism: chiral fermions in the 3-brane can be
realized as chiral fermionic zero modes of this more fundamental
theory trapped on the $\left( 3+1\right) $-dimensional defect
\cite{wchen}. In this manner the possibility that a breakdown of
general covariance on the brane due to the presence of an axial
gravitational anomaly is not ruled out, and could be induced by
quantum effects of bulk gravity.

An obvious objection to this second viewpoint is that the quantum
field theory localized on the brane will not be renormalizable. We
do not regard this objection as being fatal to the second
perspective described above. Faith in renormalization of a quantum
field theory is in part based on our ignorance of quantum gravity.
In the process of performing renormalization, ultraviolet
divergences are absorbed into redefinitions of bare parameters
such as mass and coupling constants. The presumed rationale behind
this approach to render a theory well defined is that one is
neglecting effects from quantum gravity at very short-distances,
i.e. one is just transferring ultraviolet divergences to this
regime. Hence in a model containing quantum gravity,
non-renormalizability due to gravitational effects from chiral
anomalies need not be considered as firm criteria for judging
consistency of a theory before we understand quantum gravity
completely.

\vspace{2mm}

\begin{flushleft}
{\bf 7.~Summary}
\end{flushleft}

\vspace{2mm}

We have shown that due to the presence of dynamical bulk gravity
fields and the dynamics of the 3-brane itself, the axial
gravitational anomaly depends both on both the topological
structure and the embedding of the 3-brane in the bulk. We have
found that in this brane scenario, the gravitational part of the
ABJ anomaly can be converted into an effective action for bulk
gravity without breaking local gauge symmetry on the 3-brane. It
is obvious that this effective action respects both diffeomorphism
and local Lorentz symmetries of the bulk space-time since the
induced metric is a scalar with respect to both of these
symmetries. The axial gravitational anomaly on the brane can
either imply constraints on the full bulk theory of quantum
gravity or it can be regarded as providing an observational
signature of the non-trivial vacuum structure of bulk quantum
gravity via the transfer of energy and momentum from brane to
bulk. We contend that either viewpoint is acceptable based on our
current knowledge of quantum gravity.

\acknowledgments

This work was supported by the Natural Sciences and Engineering
Research Council of Canada. We would like to thank Profs. M.
Chaichian, C. Montonen and Drs. A. Kobakhidze, F. Leblond, D.
Polyakov for useful discussions.

\end{document}